# Enhanced all-optical switching and domain wall velocity in annealed synthetic-ferrimagnetic multilayers

Luding Wang,[1,2] Youri L. W. van Hees,[2] Reinoud Lavrijsen,[2] Weisheng Zhao,[1,a] and Bert Koopmans[2,b]

[1]*Fert Beijing Institute, School of Microelectronics, Beihang University, 100191 Beijing, China*

[2]*Department of Applied Physics, Eindhoven University of Technology, P.O. Box 513, 5600 MB Eindhoven, The Netherland*

(Dated: 08 May 2020)

All-optical switching (AOS) of the magnetization in synthetic ferrimagnetic Pt/Co/Gd stacks has received considerable interest due to its high potential towards integration with spintronic devices, such as magnetic tunnel junctions (MTJs), to enable ultrafast memory applications. Post-annealing is an essential process in the MTJ fabrication to obtain optimized tunnel magnetoresistance (TMR) ratio. However, with integrating AOS with an MTJ in prospect, the annealing effects on single-pulse AOS and domain wall (DW) dynamics in the Pt/Co/Gd stacks haven't been systematically investigated yet. In this study, we experimentally explore the annealing effect on AOS and field-induced DW motion in Pt/Co/Gd stacks. The results show that the threshold fluence ($F_0$) for AOS is reduced significantly as a function of annealing temperature ($T_a$) ranging from 100°C to 300°C. Specifically, a 28% reduction of $F_0$ can be observed upon annealing at 300°C, which is a critical $T_a$ for MTJ fabrication. Lastly, we also demonstrate a significant increase of the DW velocity in the creep regime upon annealing, which is attributed to annealing-induced Co/Gd interface intermixing. Our findings show that annealed Pt/Co/Gd system facilitates ultrafast and energy-efficient AOS, as well as enhanced DW velocity, which is highly suitable towards opto-spintronic memory applications.

[a] Electronic mail: weisheng.zhao@buaa.edu.cn
[b] b.koopmans@tue.nl

The emerging potential of integrating all-optical switching (AOS) of magnetization with spintronics devices for ultrafast and energy-efficient memory applications has been soon recognized after the first observation in ferrimagnetic GdFeCo alloys,[1-3] such as employing it as part of the free layer of magnetic tunnel junctions (MTJs).[4,5] Recently, it was demonstrated that efficient single-pulse AOS can be also realized in Pt/Co/Gd synthetic ferrimagnetic multilayers with perpendicular magnetic anisotropy (PMA), in which a proximity induced atomically thin ferromagnetic region in the Gd couples antiferromagnetically to the Co.[6,7] The AOS speed of both Pt/Co/Gd stacks and GdFeCo alloys is on the (sub-)ps time scale, which is 2 orders of magnitude faster than conventional switching mechanisms in MTJs that usually operate in sub-ns regime.[8-10] However, compared with GdFeCo alloys, Pt/Co/Gd multilayers show the flexibility in fabrication and interface engineering, as well as efficient field-driven domain wall (DW) velocity[11] and considerable built-in interfacial Dzyaloshinskii-Moriya interaction (iDMI).[12] These interface-induced phenomena inherent to multilayered system are essential components for future ultra-high density memory applications.[13-16] Moreover, Co in direct contact with the MgO will increase the tunneling spin polarization (TSP), leading to a high tunnel magnetoresistance (TMR). Consequently, integrating an all-optically switchable Pt/Co/Gd stack with an MTJ has a high potential for ultrafast and ultra-high density opto-spintronic device application.[11,17-19]

Typically, as to fabrication of MTJs with high TMR and thermal stability, it is well known that thermal annealing around 300°C after film deposition is an essential process.[8,20] However, with integrating Pt/Co/Gd with an MTJ in prospect, the effects of annealing on AOS and DW dynamics has not been studied yet. Although one might expect that annealing degrades the interface, and thereby reduce interface related properties like efficient AOS and DW motion as well, very recently, Beens. et al.[21] theoretically predicted that Gd/Co interface intermixing strongly reduces the AOS threshold fluence based on the extended microscopic three-temperature model (M3TM).[22] Clearly, a systematic study of annealing effects is of utmost importance to further substantiate the applicability of the integration.

In this work, we experimentally demonstrate that single-pulse AOS is observed in Pt/Co/Gd stacks annealed up to 300°C. Moreover, annealing leads to a significant reduction of the switching threshold fluence. In addition, we show a strong increase of DW velocity in the creep regime upon increasing annealing temperature ($T_a$), while the iDMI constant reduces slightly.

The measurements are performed on Ta (4) /Pt (4)/Co (1)/ Gd (3)/Pt (2) stacks (thickness in nanometer), which are deposited on Si:B substrate at room temperature (RT) using dc magnetron sputtering at $10^{-8}$ mbar base pressure. After deposition, the thin film stacks are annealed for 0.5h with $T_a$ ranging from 0°C to 400°C. Afterwards, the magnetization of the Pt/Co/Gd stacks are measured by a VSM-SQUID at RT, as a function of out-of-plane and in-plane applied magnetic field, respectively.

Figure 1(a) shows the out-of-plane magnetic hysteresis loops of the Pt/Co/Gd stacks annealed at $T_a$ ranging from 0°C to 400°C. All these samples result in PMA, as indicated by the 100% remanence and the squareness of the hysteresis loops. Note that the coercive field ($H_c$) increases from 12.5 mT to 24 mT as $T_a$ increases, as shown in Fig. 1(b). The enhanced $H_c$ is consistent with previous studies on Co/Gd multilayers.[23] In addition, the saturation magnetization ($M_s$) shows a significant reduction upon annealing, as shown in Fig. 1(b), which is explained further on.

To investigate the annealing effects on PMA, we plot the anisotropy field ($H_k$) and the effective anisotropy ($K_{eff}$) as a function of $T_a$ in Fig. 1(c), where $H_k$ is determined from the hard-axis loops, and is calculated from $\frac{1}{2}\mu_0 M_s H_k$. We observe that $H_k$ and $K_{eff}$ show a gradual reduction as $T_a$ increases. Moreover, annealing at 400°C results in poor PMA. These results indicate that annealing leads to a significant degradation of the PMA in the Pt/Co/Gd stacks, which is consistent with previous studies reporting that annealing deteriorates PMA in a Pt/Co system.[23,24]

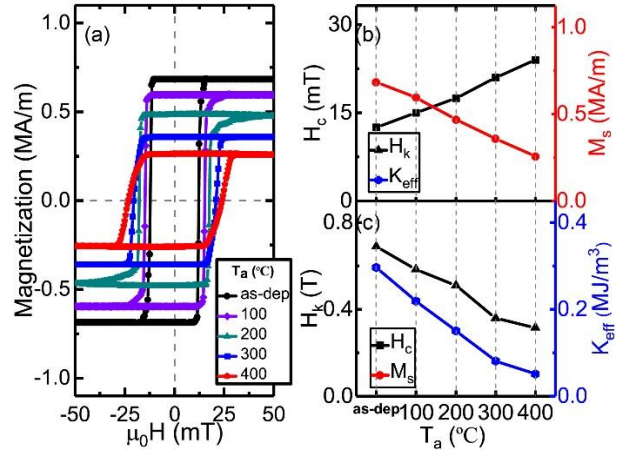

FIG. 1. (a) Hysteresis loops with out-of-plane magnetic field for the annealed Pt/Co/Gd stacks. (b) Coercive field (black cubes) and saturation magnetization (red dots) as a function of $T_a$. (c) Anisotropy field (black triangles) and effective anisotropy (blue hexagons) as a function of $T_a$.

Next, to further verify the AF exchange interaction between the Co layer and the Gd layer of the annealed Pt/Co/Gd stacks, the moment per unit area is measured as a function of temperature from 350 K to 5 K by using VSM-SQUID. As shown in Fig. 2, all the samples show AF coupling by the presence of a compensation temperature ($T_{comp}$), at which the magnetic moment of Co ($m_{Co}$) and the magnetic moment of Gd ($m_{Gd}$) compensate each other. Above $T_{comp}$, the net magnetic moment equals to $m_{Co} - m_{Gd}$. Indeed, $T_{comp}$ shows an increasing trend from 120 K to 200 K upon higher $T_a$, as shown in the inset of Fig. 2. The increased $T_{comp}$ is also in agreement with previous studies on Co/Gd multilayers, which may be resulting from annealing-induced interface related phenomena that cause modification of the exchange interaction and spin disorder/reorientation at Co/Gd interface[23].

Notably, a gradual reduction of $M_s$ at RT is also observed in Fig. 2, consistent with M(H) loops in Fig. 1(b). In the synthetic-ferrimagnetic Pt/Co/Gd, although the Curie temperature for bulk Gd is slightly lower than RT, it is elevated at the Co interface due to the presence of the antiferromagnetic (AF) coupling.[6] Upon annealing, the enhancement of interface intermixing and alloying may lead to modification of magnetic properties at the Co/Gd interface.[23] To provide a possible explanation of the reduction, the equivalent thickness of the fully magnetized Gd layer ($t_{Magn.\ Gd}$) at RT is estimated by using the data in Fig. 2. The estimation, presented in Supplementary Information Note I[25], shows that a larger $t_{Magn.\ Gd}$ is formed upon annealing, compared with a thickness of 0.43 nm for the as-dep sample (in reasonable agreement with our previous work[6]). The increased $t_{Magn.\ Gd}$ leads to a decrease of the net $M_s$ at RT. However, other interface related phenomena, such as enhanced alloying, modified alloying composition and increase of interface roughness, are also crucial and nontrival factors leading to the variation of magnetic properties, including $M_s$, in Pt/Co/Gd system. These joint mechanisms result in the annealing-induced reduction of $M_s$ observed in our experiment.

We then investigate the annealing effect on single-pulse AOS in Pt/Co/Gd stacks. In the measurements, the samples are first saturated by an external magnetic field. Afterwards, they are exposed to a number of consecutive laser pulses ranging from 1 to 5 pulses with a single-pulse energy of 400 nJ. The laser pulse is linearly polarized, with a pulse duration of ≈100 fs, a spot size of typically 100 μm and a wavelength of 700 nm. The responses of magnetization after laser excitation are measured by magneto-optical Kerr microscopy, where light and dark regions are corresponding to up and down magnetization direction.

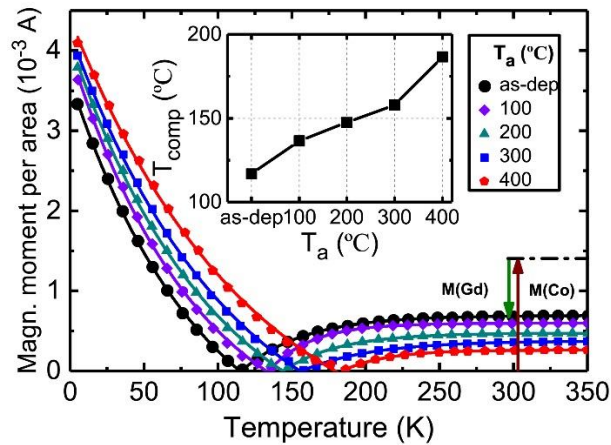

FIG. 2. VSM-SQUID measurements of the magnetic moment per unit area as a function of $T_a$ for the annealed Pt/Co/Gd stacks. Inset: compensation temperature ($T_{comp}$) as a function of $T_a$.

Figure 3(a) shows the typical AOS measurements performed on samples annealed at 300°C (see Supplemental Information Note II[25] for additional measurements on different $T_a$). For samples annealed below 400°C, a homogeneous domain with an opposite magnetization direction is written for every odd number of laser pulses, whereas no net magnetization reversal is observed for every even number of pulses. This behavior is consistent with the thermal single-pulse AOS mechanism discussed in previous studies,[7,18] in which the laser induced switching is driven by transferring the angular momentum mediated by exchange scattering between the Gd and the Co sublattices.[7]

In contrast, upon annealing at 400°C, only a multidomain state is created after any arbitrary number of pulses (as shown in Fig. 3b). These results indicate that the critical $T_a$ for AOS in Pt/Co/Gd stacks must be in the range of 300°C - 400°C. Such a high $T_a$ tolerance up to at least 300°C is promising for fabricating integrated opto-spintronic MTJ devices, since a post-annealing process at 300°C is generally required to improve TMR and thermal stability of the MTJ.[8,20]

To investigate the annealing dependence on the AOS energy efficiency, we systematically investigate the threshold fluence as a function of $T_a$ in the Pt/Co/Gd stacks. In the measurements, the samples are exposed to single laser pulses with different laser energies. Afterwards, the pulse energy dependence of the AOS written domain size is measured by Kerr microscopy (see Supplemental Information Note III[25]) and plotted in Fig. 3(c). In case of annealing below 400°C, the threshold laser energy ($P_0$) of AOS in the measurements decreases gradually upon annealing. Specifically, compared to $P_0 = 270$ nJ for the as-dep sample, it reduces to 150 nJ upon annealing at 300°C. The threshold fluence for AOS ($F_0$) can be extracted by assuming a Gaussian shape of the laser pulse (inset of Fig. 3(c)). A significant annealing dependence of $F_0$ is observed, decreasing for higher $T_a$. Note that in the case of annealing at 400°C, although no homogenous AOS domain is created for all pulse energies, the threshold fluence for the multidomain state still decreases, as shown by the open dots in Fig. 3(c).

We also observe that a minimum $F_0$ for homogenous AOS of 1.09 mJ/cm$^2$ is achieved upon annealing at 300°C, 28% lower than the $F_0$ found for as-dep sample. Moreover, the $F_0$ for annealed Pt/Co/Gd stacks is significantly lower than that for GdFeCo alloys.[1] This adds to the earlier discussed advantage of Pt/Co/Gd stacks of being more suitable for integrating with spintronics as it allows for interface engineering to be used.

The above results are consistent with recent theoretical work by Beens. et al.[21] using the extended microscopic three-temperature model (M3TM)[22], where it was demonstrated that Co/Gd interface intermixing leads to a significant reduction of $F_0$ in synthetic ferrimagnetic multilayers. This effect can be traced down to a combination of more efficient exchange scattering and reduction of the thin film's Curie temperature. In our present work, annealing leads to additional Co/Gd intermixing, which results in a decrease of $F_0$ with higher $T_a$.

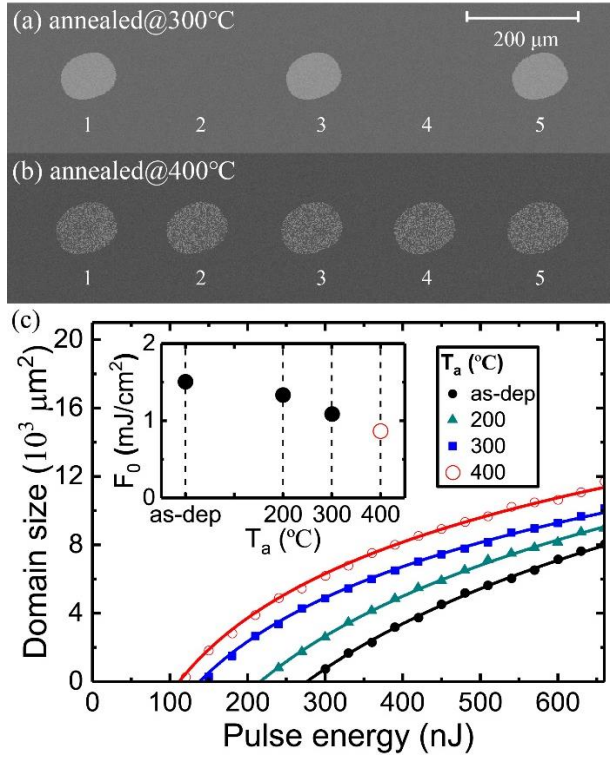

FIG. 3. (a)-(b) AOS toggle switching measurements performed on annealed Pt/Co/Gd stacks annealing 300°C and 400°C, respectively. (c) AOS domain size as a function of laser pulse energy for the annealed Pt/Co/Gd stacks. Inset: AOS threshold fluence ($F_0$) as a function of annealing temperature ($T_a$) for Pt/Co/Gd stacks.

After having discussed effects on AOS, we change our attention to annealing effects on DW dynamics and DMI in Pt/Co/Gd stacks. Previous studies[6,12] reported that Pt/Co/Gd stacks exhibit efficient field-driven DW motion and a considerable DMI value, which are highly promising for integrating AOS with DW devices[26-28] or skyrmion-electronics.[13,29] Figure 4(a) shows a typical measurement of field-induced DW motion. The velocity is extracted from the average expansion of a bubble domain under an out-of-plane magnetic field pulse ($\mu_0 H_z$) with a pulse duration ranging from 10 μs to 1000 μs. The breaking of radial symmetry in the measurement originates from the presence of an effective iDMI field. In Fig. 4(b), the DW velocity as a function of $\mu_0 H_z$ for all $T_a$ is plotted on a logarithmic scale according to the universal creep law[30],

$$v = v(H_{dep}) \exp\left[-\frac{U_c}{k_B T}\left(\left(\frac{H_{dep}}{H_z}\right)^\mu - 1\right)\right] \quad (1)$$

where $U_c$ is the scaling energy constant, $k_B$ is the Boltzmann constant, $T$ is the temperature, $H_{dep}$ is the depinning field at 0 K, and $\mu = 1/4$ is the critical exponent for a 2D system (see Supplemental Information Note IV[25]). Remarkably, a significant enhancement of DW velocity in the creep regime is observed upon increasing $T_a$. Note that the DW velocity in the creep regime is mainly determined by the pinning of DWs, which originates from structural defects in the ultrathin film system. To further explain the enhancement of DW velocity, the pinning potential ($P$)

$$P = \left(\frac{U_c}{k_B T}\right) H_{dep} \qquad (2)$$

as a function of $T_a$ is plotted in Fig. 4(c), which is correlated with the linear slope in Fig. 4(b). The reduction of $P$ upon annealing indicates a progressive decrease of DW pinning, thus causing the enhancement of the DW velocity in the creep regime.

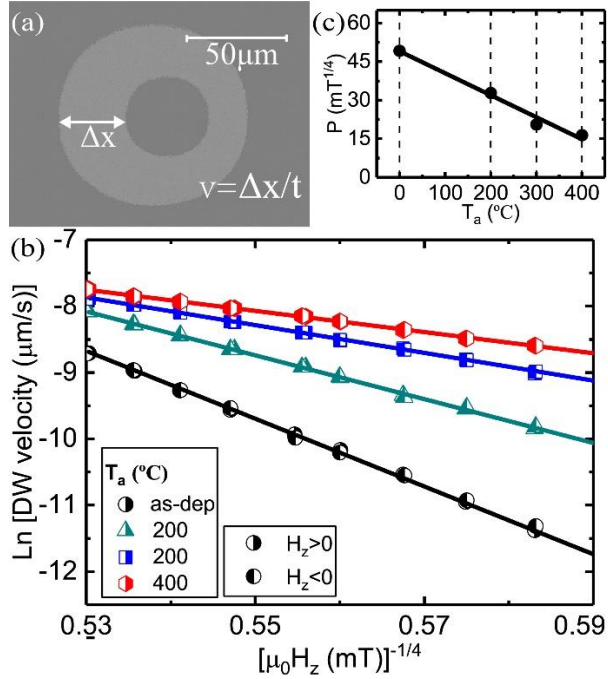

FIG. 4. (a) Typical differential Kerr image for DW velocity measurement for annealed Pt/Co/Gd stack (b) DW velocity as a function of out-of-plane magnetic field pulse. (c) Pinning potential ($P$) as a function of annealing temperature ($T_a$) for Pt/Co/Gd stacks.

Finally, to investigate the variation of iDMI in Pt/Co/Gd stacks upon annealing, the DMI effective field ($\mu_0 H_{DMI}$) is measured by asymmetric bubble-domain expansion under $\mu_0 H_z$ as a function of in-plane magnetic field ($\mu_0 H_x$). Figure 5(a)-(b) show the typical differential Kerr images under an in-plane field of ±340 mT. Notably, the asymmetry of DW expansion along the x axis in the measurements is due to the interplay between the internal $\mu_0 H_{DMI}$ and the external $\mu_0 H_x$, from which $\mu_0 H_{DMI}$ can be estimated by $\mu_0 H_x$ corresponding to the slowest DW speed (see Supplemental Information Note V[25]). Figure 5(c) shows $\mu_0 H_{DMI}$ as a function of $T_a$ in Pt/Co/Gd stacks, where a slight decrease is observed upon annealing. Specifically, $\mu_0 H_{DMI}$ for the as-dep sample is around 300 mT, whereas in case of $T_a = 300°C$, $\mu_0 H_{DMI}$ reduces to 250 mT. The DMI constant ($D$) is then deduced from $\mu_0 H_{DMI} = D/M_s\sqrt{A/K_{eff}}$, where A is the exchange stiffness constant of 16 pJ/m taken from literature and assumed to be constant for all samples [12]. Note that a DMI constant of 1.1 mJ/m$^2$ is observed for the as-dep Pt/Co/Gd stack, which is in reasonable agreement with previous report[12]. In addition, although D shows a decreasing trend with $\mu_0 H_{DMI}$ when increasing of $T_a$ (shown in Fig. 5(c)), a considerable D value of 0.75 mJ/m$^2$ is still observed for 300°C annealing. Like the trends in AOS and DW velocities, the results on iDMI could be explained by the modification of the magnetic properties induced by intermixing, alloying and other related mechanisms at the Pt/Co and Co/Gd interfaces, which leads to subtle changes of the exchange interaction between the atoms near interfaces.

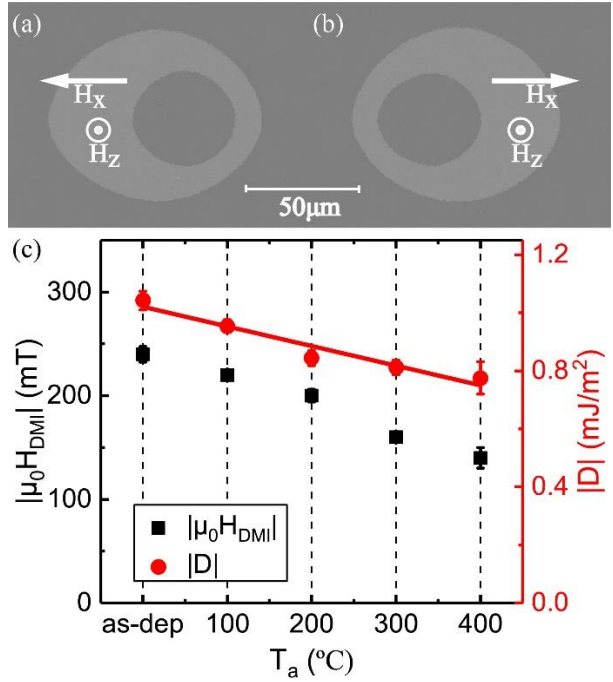

FIG. 5. (a)-(b). Typical differential Kerr images for iDMI measurement for the annealed Pt/Co/Gd stacks. (c) effective DMI field (black cubes) and DMI constant (red dots) of the annealed Pt/Co/Gd stacks as a function of $T_a$.

In conclusion, we have experimentally demonstrated enhancing AOS and DW velocity through annealing in synthetic ferrimagnetic Pt/Co/Gd stacks. Notably, toggle switching behavior is observed upon annealing up to 300°C, which is essential for the integration with MTJ devices due to the required post-annealing process. Moreover, the AOS threshold fluence reduces significantly upon annealing, which indicates an enhancement of AOS energy efficiency for data writing. In addition, a significantly enhancement of DW velocity upon annealing is also observed, which is in accordance with a reduction of the DW pinning potential at the Co/Gd interface. Lastly, a considerable DMI value is observed for samples annealed up to 300°C. Therefore, our results demonstrate that the synthetic ferrimagnetic Pt/Co/Gd stack is an ideal candidate for integrating AOS in spintronic devices, such as MTJs or magnetic DW devices, which may pave the way towards ultrafast and energy efficient opto-spintronics memory application.

This work is part of the Gravitation program 'Research Centre for Integrated Nanophotonics', which is financed by the Netherlands Organisation for Scientific Research (NWO).

We gratefully acknowledge the National Key Technology Program of China 2017ZX01032101 and National Natural Science Foundation of China Grant No. 61627813, the International Collaboration Project B16001 and the Beihang Hefei Innovation Research Institute Project BHKX-17-06 and China Scholarship Council (CSC) for their financial support of this work.

[1] C. D. Stanciu, F. Hansteen, A. V. Kimel, A. Kirilyuk, A. Tsukamoto, A. Itoh, and Th. Rasing, Phys. Rev. Lett. 99, 047601 (2007).

[2] T. A. Ostler, J. Barker, R. F. L. Evans, R. W. Chantrell, U. Atxitia, O. Chubykalo-Fesenko, S. El Moussaoui, L. Le Guyader, E. Mengotti, L. J. Heyderman, F. Nolting, A. Tsukamoto, A. Itoh, D. Afanasiev, B. A. Ivanov, A. M. Kalashnikova, K. Vahaplar, J. Mentink, A. Kirilyuk, Th. Rasing, and A. V. Kimel, Nat. Commun. 3, 666 (2012).

[3] I. Radu, K. Vahaplar, C. Stamm, T. Kachel, N. Pontius, H. A. Durr, T. A. Ostler, J. Barker, R. F. L. Evans, R. W. Chantrell, A. Tsukamoto, A. Itoh, A. Kirilyuk, Th. Rasing, and A. V. Kimel, Nature (London) 472, 205 (2011).

[4] J. Chen, L. He, J. Wang, and M. Li, Phys. Rev. Appl. 7, 021001 (2017).

[5] L. Avilés-Félix, L. Álvaro-Gómez, G. Li, C. S. Davies, A. Olivier, M. Rubio-Roy, S. Auffret, A. Kirilyuk, A. V. Kimel, Th. Rasing, L. D. Buda-Prejbeanu, R. C. Sousa, B. Dieny, and I. L. Prejbeanu, AIP Adv. 9, 125328 (2019).

[6] M. L. M. Lalieu, M. J. G. Peeters, S. R. R. Haenen, R. Lavrijsen, and B. Koopmans, Phys. Rev. B 96, 220411 (2017).

[7] M. Beens, M. L. M. Lalieu, A. J. M. Deenen, R. A. Duine, and B. Koopmans, Phys. Rev. B 100, 220409 (2019).

[8] M. Wang, W. Cai, K. Cao, J. Zhou, J. Wrona, S. Peng, H. Yang, J. Wei, W. Kang, Y. Zhang, J. Langer, B. Ocker, A. Fert, and W. Zhao, Nat. Commun. 9, 617 (2018).

[9] S. Peng, D. Zhu, J. Zhou, B. Zhang, A. Cao, M. Wang, W. Cai, K. Cao, and W. Zhao, Adv. Electron. Mater. 5, 1900134 (2019).

[10] M. Wang, W. Cai, D. Zhu, Z. Wang, J. Kan, Z. Zhao, K. Cao, Z. Wang, Y. Zhang, T. Zhang, C. Park, J.-P. Wang, A. Fert, and W. Zhao, Nat. Electron. 1, 582 (2018).

[11] M. L. M. Lalieu, R. Lavrijsen, and B. Koopmans, Nat. Commun. 10, 110 (2019).

[12] T. H. Pham, J. Vogel, J. Sampaio, M. Vaňatka, J.-C. Rojas-Sánchez, M. Bonfim, D. S. Chaves, F. Choueikani, P. Ohresser, E. Otero, A. Thiaville, and S. Pizzini, Europhys. Lett. 113, 67001 (2016).

[13] J. Sampaio, V. Cros, S. Rohart, A. Thiaville, and A. Fert, Nat. Nanotechnol. 8, 839 (2013).

[14] N. Penthorn, X. Hao, Z. Wang, Y. Huai, and H. Jiang, Phys. Rev. Lett. 122, 257201 (2019).

[15] C. Hanneken, F. Otte, A. Kubetzka, B. Dupé, N. Romming, K. vonBergmann, R. Wiesendanger, and S. Heinze, Nat. Nanotechnol. 10, 1039 (2015).

[16] S.-G. Je, P. Vallobra, T. Srivastava, J.-C. Rojas-Sánchez, T. H. Pham, M. Hehn, G. Malinowski, C. Baraduc, S. Auffret, G. Gaudin, S. Mangin, H. Béa, and O. Boulle, Nano Lett. 18, 7362 (2018).

[17] A. V. Kimel and M. Li, Nat. Rev. Mater. 4, 189 (2019).

[18] S. Mangin, M. Gottwald, C-H. Lambert, D. Steil, V. Uhlíř, L. Pang, M. Hehn, S. Alebrand, M. Cinchetti, G. Malinowski, Y. Fainman, M. Aeschli-mann, and E. E. Fullerton, Nat. Mater. 13, 286 (2014).

[19] P. Liu, X. Lin, Y. Xu, B. Zhang, Z. Si, K. Cao, J. Wei, and W. Zhao, Materials (Basel) 11, 47 (2017).

[20] W. Zhao, X. Zhao, B. Zhang, K. Cao, L. Wang, W. Kang, Q. Shi, M. Wang, Y. Zhang, Y. Wang, S. Peng, J.-O. Klein, L. de Barros Naviner, and D. Rav-elosona, Materials (Basel) 9, 41 (2016).

[21] M. Beens, M. L. M. Lalieu, R. A. Duine, and B. Koopmans, AIP Adv. 12, 125133 (2019).

[22] B. Koopmans, G. Malinowski, F. Dalla Longa, D. Steiauf, M. Fähnle, T. Roth, M. Cinchetti, and M. Aeschlimann, Nat. Mater. 9, 259 (2010).

[23] M. A. Basha, C. L. Prajapat, M. Gupta, H. Bhatt, Y. Kumar, S. K. Ghosh, V. Karki, S. Basu, and S. Singh, Phys. Chem. Chem. Phys. 20, 21580 (2018).

[24] T. Young Lee, D. Su Son, S. Ho Lim, and S.-R. Lee, Appl. Phys. Lett 113, 216102 (2013).

[25] See Supplemental Information for (I) analysis of the SQUID M(T) measurements on the anealed Pt/Co/Gd stack; (II III) measurements performed on samples annealed at different $T_a$ for up to 10 subsequent laser pulses; (III) pulse-energy-dependent AOS measurement on a Pt/Co/Gd stack with different $T_a$; (IV) DW velocity as a function of $H_z$ plotted on normal scale; and (V) HDMI measurements performed on annealed Pt/Co/Gd stacks.

[26] B. Zhang, Y. Xu, W. Zhao, D. Zhu, X. Lin, M. Hehn, G. Malinowski, D. Ravelosona, and S. Mangin, Phys. Rev. Appl. 11, 034001 (2019).

[27] B. Zhang, Y. Xu, W. Zhao, D. Zhu, H. Yang, X. Lin, M. Hehn, G. Ma-linowski, N. Vernier, D. Ravelosona, and S. Mangin, Phys. Rev. B 99, 144402 (2019).

[28] Y. Xu, M. Deb, G. Malinowski, M. Hehn, W. Zhao, and S. Mangin, Adv. Mater. 29, 1703474 (2017).

[29] W. Kang, Y. Huang, X. Zhang, Y. Zhou, and W. Zhao, Proc. IEEE 104, 2040 (2016).

[30] V. Jeudy, A. Mougin, S. Bustingorry, W. Savero Torres, J. Gorchon, A. B. Kolton, A. Lemaître, and J.-P. Jamet, Phys. Rev. Lett. 117, 057201 (2016)

# Supplementary Information

**Enhanced all-optical switching and domain wall velocity in annealed synthetic-ferrimagnetic multilayers**

Luding Wang,[1, 2] Youri L.W. van Hees,[2] Reinoud Lavrijsen,[2] Weisheng Zhao,[1,a)]

and Bert Koopmans[2,b)]

[1] *Fert Beijing Institute, School of Microelectronics, Beihang University, 100191 Beijing, China*

[2] *Department of Applied Physics, Eindhoven University of Technology, P.O. Box 513, 5600 MB Eindhoven, The*

*Netherlands*

(Dated: 08 May 2020)

---

a) weisheng.zhao@buaa.edu.cn
b) b.koopmans@tue.nl

# I. Analysis of the SQUID-VSM measurement on the annealed Pt/Co/Gd stack

| $T_a$ (°C) | $t_{Magn.Gd}$ (nm) |
|---|---|
| 0 | 0.43 |
| 100 | 0.45 |
| 200 | 0.51 |
| 300 | 0.56 |
| 400 | 0.59 |

Table S1. The thickness of proximity induced fully magnetized Gd layer at room temperature ($t_{Magn.Gd}$) as a function of $T_a$. An enhancement of $t_{Magn.Gd}$ upon higher $T_a$ indicating a possible explanation for the reduction of $M_s$.

To quantitatively analyze the annealing effect on the reduction of $M_s$ in Pt/Co/Gd stacks from a synthetic ferrimagnet viewpoint, the equivalent thickness of the fully magnetized Gd layer as a function of $T_a$ at room temperature can be extracted from the data in the Fig. 2 of the main text.

The detailed estimation process is as follows, which is the same method as used in a previous study[1]. In case of $T_a$ = 300°C, a magnetic moment per unit area of 4.22 MA/m at 0 K can be extrapolated from Fig. 2. By using the magnetization of 1.4 MA/m for bulk Co, 1 nm Co layer results in the magnetic moment per unit area of 1.4 Am (as indicated as red arrow in Fig. 2). Due to the antiferromagnetic coupling of the Co and Gd sublattices, this leaves a magnetic moment per unit area of 5.62 Am for the Gd film. This value corresponds to a magnetization of 1.87 MA/m for 3 nm Gd layer, which is consistent with the magnetization of bulk Gd. At room temperature (300 K), a magnetic moment per unit area of 0.36 Am is measured. Above $T_{comp}$, this value equals to the moment of the Co minus that of the Gd. By using the moment per unit area of 1 nm Co layer, a magnetic moment per unit area of 1.04 Am for the Gd layer can be obtained. Using the magnetization for the Gd layer, the value of 1.04 Am corresponds to 0.56 nm of fully saturated Gd, which is in the same order as found in previous studies.

Based on this method, the thickness of fully magnetized Gd layer at room temperature ($t_{Magn.Gd}$) as a function of $T_a$ can be determined and is shown in Table S1. We observe an enhancement of $t_{Magn.Gd}$ upon higher $T_a$. More specifically, compared with 0.43 nm for as-dep sample, $t_{Magn.Gd}$ increased to 0.56 nm in case of $T_a$ =300°C. Due to the antiferromagnetic coupling between the Co and the Gd layer, the increased $t_{Magn.Gd}$ leads to a reduction of net moment per unit area at room temperature.

This estimation demonstrates a possible explanation for the reduction of $M_s$ as found in our experiment on annealed Pt/Co/Gd stacks. This result is also consistent with previous studies on Gd/Co multilayers[2]. However, it is worth noting that other annealing induced interface phenomena, such as alloying and increase of interface roughness at Co/Gd interface are also crucial factors during annealing. Under the joint effects of all these annealing-induced interface related mechanisms, the modification of magnetic properties in Pt/Co/Gd stacks eventually result in the reduction of $M_s$ observed in our experiment.

## II. Single-pulse toggle AOS measurement for the annealed Pt/Co/Gd stacks

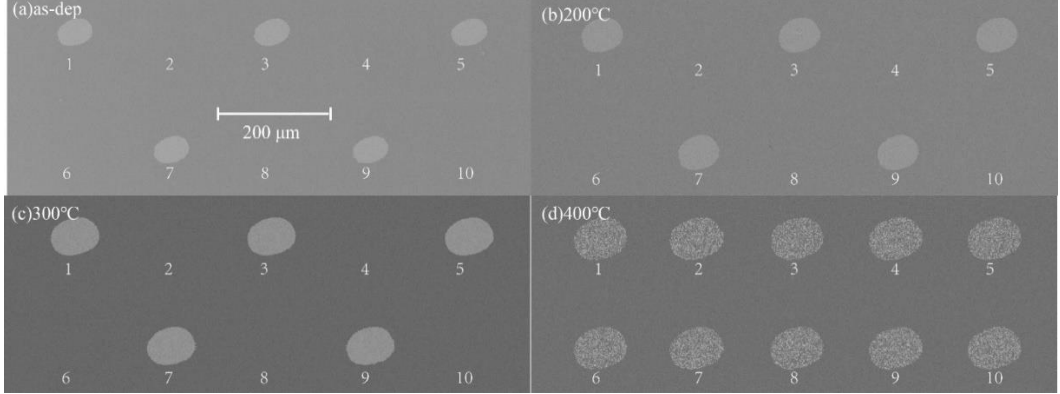

FIG. S2. AOS toggle switching measurements performed on annealed Pt/Co/Gd stacks after annealing for 0.5 h at 0, 200°C, 300°C and 400°C, respectively. The number of laser spots are indicated in the figure, up to 10 consecutive laser pulses. We observe single pulse toggle switching upon $T_a \leq 300°C$. However, in case of $T_a = 400°C$, a multidomain state is created for any arbitrary number of pulses.

In the main text, we investigate annealing effects on AOS toggle switching for Pt/Co/Gd stacks. The consecutive laser pulses measurement up to 10 laser pulses is performed at $T_a = 0$, 200°C, 300°C, 400°C, respectively, as shown in Fig. S2.

In case of $T_a \leq 300°C$, a homogeneous domain with an opposite magnetization direction is written for every odd number of laser pulses, whereas no net magnetization reversal is observed for every even number of pulses. These results demonstrate that each next laser pulse completely reverses the domain created by a previous sequence of AOS events. The observed switching behavior is consistent with previous studies on the thermal single-pulse switching mechanism.

In case of $T_a = 400°C$, the absence of single-pulse AOS demonstrates that exceeding an annealing temperature higher than 300°C leads to the loss of single-pulse AOS behavior in Pt/Co/Gd, which is discussed in more detail in the main text. These results indicate that the threshold $T_a$ for single-pulse AOS in Pt/Co/Gd stacks is above 300°C, which is a crucial temperature for MTJ post-annealing process.

**III. Pulse-energy dependent AOS measurement on the annealed Pt/Co/Gd stack**

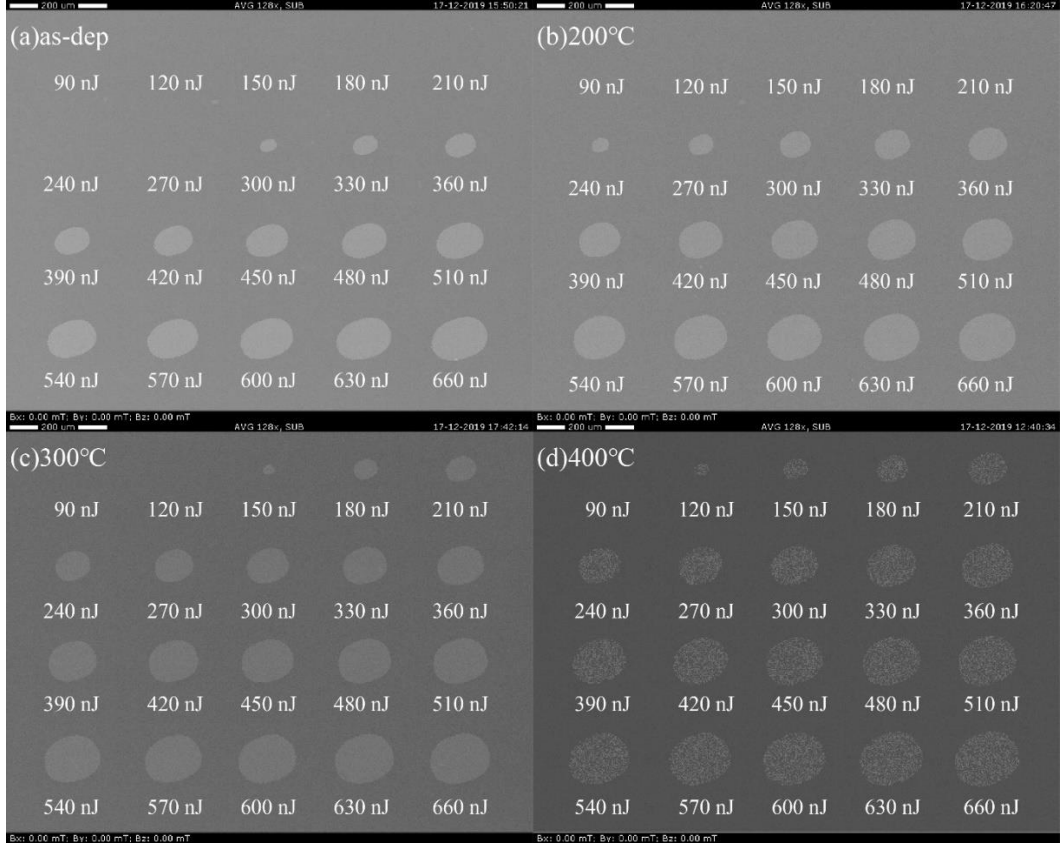

FIG. S3. Kerr microscope image of the pulse-energy dependence of domain size measurements performed on the Pt/Co/Gd stacks annealed at different $T_a$. Laser pulse energies are indicated in the figure. The results show that in case of $T_a \leq 300°C$, a decreasing threshold laser energy is observed upon higher $T_a$, while in case of $T_a = 400°C$, a multidomain state is created for all pulse energies from the threshold laser energy.

In the main text, we investigate annealing effects on AOS threshold fluence for Pt/Co/Gd stacks. The threshold fluence is determined by pulse-energy dependent measurements, as shown in Fig. S3. The size of domains and its dependence on laser energy is calculated and analyzed in Fig. 3(c) of the main text.

Figure S3(a)-(c) show that, in case of $T_a \leq 300°C$, a homogenous domain is created after the laser excitation, as found in previous work. Moreover, the threshold laser energy $P_0$ decreases from 300 nJ for as-dep sample to 150nJ upon $T_a = 300°C$. These results show that the energy efficiency of AOS in Pt/Co/Gd stacks is enhanced upon annealing in the range of 0 - 300°C.

However, in case of $T_a = 400°C$, a multidomain state can be written from $P_0$ at 120 nJ and up, showing that exceeding $T_a$ can lead to substantial change of AOS inherent to annealed Pt/Co/Gd stacks (discussed in the main text).

## IV. DW velocity as a function of magnetic field pulse of the annealed Pt/Co/Gd stacks

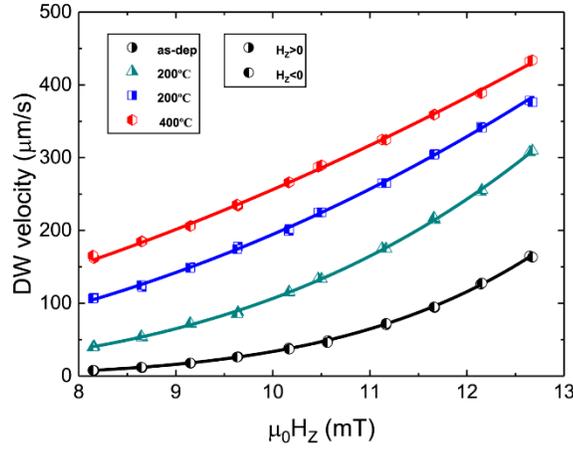

FIG. S4 Domain wall velocity as a function of out-of-plane magnetic field pulse ($\mu_0 H_z$) for Pt/Co/Gd stacks with different $T_a$ (plotted in normal scale).

To investigate annealing effects on field-driven DW velocity in the creep regime, Fig. S4 plots the DW velocity on a linear scale as a function of out-of-plane magnetic field pulse ($\mu_0 H_z$) for both directions ($\mu_0 H_z < 0$ and $\mu_0 H_z > 0$, respectively), for Pt/Co/Gd stacks annealed at 0, 200°C, 300°C, 400°C, respectively. In the figure, a significant enhancement of DW velocity in the creep regime upon annealing can be observed directly.

## V. Determination of the DMI effective field for annealed Pt/Co/Gd stacks

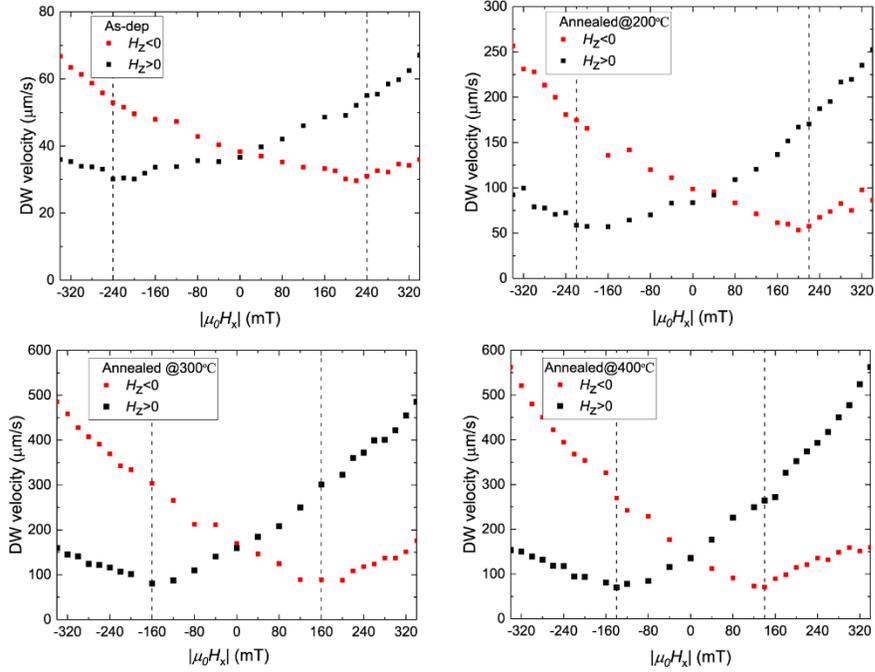

FIG. S5 Domain wall velocity under out-of-plane magnetic field pulse ($\mu_0 H_z$) as a function of in-plane magnetic field ($\mu_0 H_x$) for Pt/Co/Gd stacks annealed at 0, 200°C, 300°C, 400°C, respectively.

To investigate the variation of the DMI effective field ($\mu_0 H_{DMI}$) upon annealing, asymmetric bubble-domain expansion measurements are performed on Pt/Co/Gd annealed stacks.

In the measurements, the DW motion is driven by an out-of-plane magnetic field pulse $\mu_0 H_z$ with both directions (as indicated in red and black cubes for $\mu_0 H_z < 0$ and $\mu_0 H_z > 0$, respectively) as a function of in-plane magnetic field ($\mu_0 H_x$). Since the asymmetry of DW expansion along the x axis in the measurements is due to the interplay between the internal $\mu_0 H_{DMI}$ and the external $\mu_0 H_x$, the effective DMI field can be estimated by $\mu_0 H_x$ corresponding to the slowest DW speed[3].

Figure S5 shows the result of the asymmetric bubble-domain expansion measurement performed on Pt/Co/Gd stack annealed at 0, 200°C, 300°C, 400°C, respectively. Examining the domain wall velocity as a function of the $\mu_0 H_x$, the DMI effective field corresponding to the slowest DW velocity is determined by fitting the parabola, as indicated by the vertical dashed lines in the figure. These results constitute Fig. 5(c) in the main text.

**Supplementary reference**

[1]M. L. M. Lalieu, M. J. G. Peeters, S. R. R. Haenen, R. Lavrijsen, and B.Koopmans, Phys. Rev. B96, 220411 (2017).

[2] M. A. Basha, C. L. Prajapat, M. Gupta, H. Bhatt, Y. Kumar, S. K. Ghosh, V. Karki, S. Basu, and S. Singh, Phys. Chem. Chem. Phys.20, 21580(2018).

[3]T. H. Pham, J. Vogel, J. Sampaio, M. Vaňatka, J.-C. Rojas-Sánchez, M.Bonfim, D. S. Chaves, F. Choueikani, P. Ohresser, E. Otero, A. Thiaville,and S. Pizzini, Europhys. Lett.113, 67001 (2016).